\documentclass[conference]{IEEEtran}
\newif\ifdraft
\drafttrue

\usepackage{cite}
\usepackage{amsmath}
\usepackage{amssymb}
\usepackage{amsfonts}
\usepackage{mathtools}
\usepackage{bbold}
\usepackage{soul}
\usepackage{booktabs}
\usepackage{float}
\usepackage{tcolorbox}
\usepackage{graphicx}
\usepackage{textcomp}
\usepackage{xcolor}
\usepackage{xspace}
\usepackage{mathrsfs}
\usepackage{hyperref}
\usepackage{url}
\usepackage[inline]{enumitem}

\usepackage{tikz}
\usetikzlibrary{arrows.meta}
\usetikzlibrary{calc, babel, decorations.text, fit}
\usetikzlibrary{positioning, intersections}
\usepgflibrary{shapes.arrows}

\usepackage{pgfplots}
\usepackage{pgfplots,pgfplotstable}

\usepackage{amsthm}
\usepackage{algorithm,algpseudocode}
\usepackage{dsfont}
\usepackage{subfig}
\usepackage{mwe}
\usepackage{wrapfig}
\usepackage{lipsum}
\usepackage{framed}
\usepackage{svg}
\usepackage{multirow}
\usepackage{soul}
\usepackage{wasysym}

\usepackage{enumitem}
\usepackage{graphicx}
\usepackage{indentfirst}
\usepackage[nottoc]{tocbibind}
\usepackage[colorinlistoftodos,disable]{todonotes}
\hypersetup{
    colorlinks=true,
    linkcolor=blue,
    filecolor=magenta,      
    urlcolor=cyan,
}

\ifdraft
\newcommand{\anvith}[1]{\textcolor{blue}{Anvith: #1}}
\newcommand{\mohammad}[1]{\textcolor{teal}{MY: #1}}
\newcommand{\nick}[1]{\textcolor{olive}{HJ: #1}}
\newcommand{\adelin}[1]{\textcolor{purple}{AT: #1}}
\newcommand{\nicolas}[1]{\textcolor{red}{Nicolas: #1}}
\newcommand{\varun}[1]{\textcolor{red}{Varun: #1}}
\newcommand{\jonas}[1]{\textcolor{orange}{Jonas: #1}}
\newcommand{\new}[1]{\textcolor{black}{#1}}
\newcommand{\roei}[1]{\textcolor{purple}{Roei: #1}}
\else
\newcommand{\anvith}[1]{}
\newcommand{\mohammad}[1]{}
\newcommand{\nick}[1]{}
\newcommand{\adelin}[1]{}
\newcommand{\nicolas}[1]{}
\newcommand{\varun}[1]{}
\newcommand{\jonas}[1]{}
\newcommand{\new}[1]{}
\newcommand{\roei}[1]{}
\fi

\newcommand{\ie}{\textit{i.e.,}\@\xspace}
\newcommand{\eg}{\textit{e.g.,}\@\xspace}
\newcommand{\etal}{\textit{et al.}\@\xspace}

\newcommand{\model}{M\xspace}
\newcommand{\weights}{\textbf{w}\xspace}

\theoremstyle{definition}

\newcommand\blfootnote[1]{%
  \begingroup
  \renewcommand\thefootnote{}\footnote{#1}%
  \addtocounter{footnote}{-1}%
  \endgroup
}

\usepackage{tikz}
\usetikzlibrary{arrows.meta}
\usetikzlibrary{calc, babel, decorations.text}
\usetikzlibrary{positioning}
\usepgflibrary{shapes.arrows}
\usetikzlibrary{patterns}
\usetikzlibrary{decorations.text}    
\usetikzlibrary{intersections}

\makeatletter
\tikzset{arc style/.initial={}}
\pgfdeclareshape{half circle}{
  \inheritsavedanchors[from=circle]
  \inheritanchorborder[from=circle]

  \inheritanchor[from=circle]{center}
  \inheritanchor[from=circle]{south}
  \inheritanchor[from=circle]{west}
  \inheritanchor[from=circle]{north}
  \inheritanchor[from=circle]{east}

  \inheritbackgroundpath[from=circle]

  \beforebackgroundpath{
    \pgfkeys{/tikz/arc style/.get=\tmp}
    \expandafter\tikzset\expandafter{\tmp}
    \tikz@options

    \radius \pgf@xa=\pgf@x
    \centerpoint \pgf@xb=\pgf@x \pgf@yb=\pgf@y

    \advance\pgf@yb by \pgf@xa
    \pgfpathmoveto{\pgfpoint{\pgf@xb}{\pgf@yb}}
    \pgfpatharc{90}{-90}{\pgf@xa}

    \pgfusepath{fill}
  }
}
\makeatother

\tikzset{
	partial ellipse/.style args={#1:#2:#3}{
		insert path={+ (#1:#3) arc (#1:#2:#3)}
	}
}

\usepackage{pstricks}

\pdfoutput=1

\begin{document}

\title{SoK: Machine Learning Governance}

\author{Varun Chandrasekaran\text{*}\IEEEauthorrefmark{2}, Hengrui Jia\text{*}\IEEEauthorrefmark{3}\IEEEauthorrefmark{4}, Anvith Thudi\text{*}\IEEEauthorrefmark{3}\IEEEauthorrefmark{4},\\ Adelin Travers\text{*}\IEEEauthorrefmark{3}\IEEEauthorrefmark{4}, Mohammad Yaghini\text{*}\IEEEauthorrefmark{3}\IEEEauthorrefmark{4}, Nicolas Papernot\IEEEauthorrefmark{3}\IEEEauthorrefmark{4}\vspace*{0.15cm} \\ 
University of Toronto\IEEEauthorrefmark{3}, Vector Institute\IEEEauthorrefmark{4}, University of Wisconsin-Madison\IEEEauthorrefmark{2} 
}

\maketitle
\thispagestyle{plain}
\pagestyle{plain}

\begin{abstract}
The application of machine learning (ML) in computer systems introduces not only many benefits but also risks to society. In this paper, we develop the concept of ML governance to balance such benefits and risks, with the aim of achieving responsible applications of ML. Our approach first systematizes research towards ascertaining ownership of data and models, thus fostering a notion of \textit{identity} specific to ML systems. Building on this foundation, we use identities to hold principals \textit{accountable} for failures of ML systems through both attribution and auditing. To  increase trust in  ML systems, we then survey techniques for developing \textit{assurance}, \ie confidence that the system meets its security requirements and does not exhibit certain known failures. This leads us to highlight the need for techniques that allow a model owner to manage the \textit{life cycle} of their system, \eg to patch or retire their ML system. Put altogether, our systematization of knowledge standardizes the interactions between principals involved in the deployment of ML throughout its life cycle. We highlight opportunities for future work, \eg to formalize the resulting game between ML principals. 
\end{abstract}
\blfootnote{\text{*}All student authors contributed equally and are ordered alphabetically.}
\section{Introduction}

Like any maturing technology, machine learning (ML) is transitioning from a stage of unfettered innovation to increasing calls to mitigate the risks associated with its deployment. In the context of ML systems, these potential risks include privacy violations (\eg leaking sensitive training data), unfair predictions (\eg in facial recognition), misinformation (\eg generative models facilitating the creation of deepfakes), or a lack of integrity (\eg self-driving car accidents due to an incorrect understanding of their environment). At the societal level, there is a need to ensure that applications of ML balance these risks with the benefits brought by their improved predictive capabilities, just as practical systems security balances  the cost of protection with the risk of loss~\cite{lampson2004computer}. 

\textit{Machine learning governance} standardizes the interactions between principals involved in the deployment of ML with the aim of ensuring a responsible application of ML, \ie one that is beneficial to society.%

Governance in ML differs from existing research in several ways. Indeed, the attack surface of ML algorithms has been the subject of many studies; for instance, work on adversarial ML~\cite{huang2011adversarial} evaluated the worst-case performance of ML under various threat models. However, this line of work often focuses on narrow aspects, such as the technical limitations of ML algorithms, and studies these limitations in isolation. For instance, while it is known how to train a model privately~\cite{chaudhuri2011differentially,abadi2016deep}, the impact of private learning on model robustness is less studied~\cite{song2019privacy} thus affecting our ability to deploy such algorithms in settings where the model must be aligned with multiple social norms. %
It, thus, remains unclear how companies and institutions applying ML should manage the different risks associated with ML. This is exacerbated in the online setting where a model is not static but rather iteratively trained and deployed. %
These difficulties have also manifested themselves in inadequate regulation, as illustrated by the focus on anonymization in privacy regulation~\cite{cummings2018role} despite its well-known limitations~\cite{narayanan2008robust}. As a result, regulators are unable to formalize, implement, and enforce regulation that contains the societal risks of ML.

In this paper, we systematize knowledge from security and privacy research on ML to outline a practical approach to ML governance. Our approach is four-fold and can be thought of as defining the types of interactions that principals (defined in  \S~\ref{sec:principals}) of an ML ecosystem will require to develop the trust needed to ensure responsible applications of ML.  While existing research on adversarial ML (and more broadly trustworthy ML) serves as a foundation for some of the approaches, our systematization identifies open problems which the computer security community needs to tackle. 

First, we observe that trust requires that one make unforgeable and undeniable claims of ownership about an ML model and its training data. This establishes the concept of \textit{identity}, which identifies a key principal in the ML application: its owner. This is a prerequisite to holding model developers accountable for the potential negative consequences of their ML algorithms: if one is unable to prove that a model belongs to a certain entity, it will be impossible to hold the entity accountable for the model's limitations. 

Once ownership of the model is established, the second component of our approach to ML governance is \textit{accountability}. Here, we attribute model failures to the model owner and its end users, and establish the responsibility of these different principals for failures of the model. These failures may have been identified through the use of the system, but could also result from external scrutiny by entities such as regulators. We thus also discuss how to audit an ML application to identify and attribute failures that have yet to be discovered. 

To prevent failures in their ML application, and to foster trust from end users, model owners will then deploy a set of techniques to obtain \textit{assurance}. Loosely defined, these techniques establish confidence that the system meets specific requirements. The model owner here wishes to make certain claims establishing guarantees of controlled risk for their ML application. While these claims are currently often made post-hoc, by inspecting the trained model, we argue that future research should focus on the training algorithm. We discuss ways that these claims can be substantiated by, for instance, proactively collecting evidence during training  or providing documentation to end users. 

Finally, the fourth component of our approach realizes that the trade-off between ML risks and benefits will be regularly re-evaluated by the different principals involved. As such, we outline how there will be a need for mechanisms that allow one to dynamically edit models after they were released: this lifecycle management is analogous to common practices in software development like versioning. This represents challenges in how one can remove behavior already learned by a model (\ie machine unlearn) and learn new behaviors, but also retire the model altogether.

These interactions between model owners, users, and regulators are at the core of our approach to ML governance and result in a formal formulation of what it means to balance the risks of applying ML (see \S~\ref{sec:game}). Concretely, we present a Stackelberg game among the principals involved in an ML governance scenario. The approach is more general, and can capture competing interests of the principals as well as the possible trade-offs, thus delineating the Pareto frontier. 

To summarize, our contributions are the following: 
\begin{itemize}
    \item We define the problem of ML governance as balancing the risks of ML with its benefits. Specifically, ML governance supposes that society has defined norms that serve as the basis for characterizing trade-offs between risks and benefits. We formalize the interactions between ML principals: \ie model owners making claims of having managed certain risks (\eg privacy, fairness, robustness, etc.), end user who seek to trust these claims, and regulators who seek to audit such claims and verify compliance.
    \item We systematize approaches to establishing identity in ML through data and model ownership in \S~\ref{sec:ownership}. This serves as the foundation for holding model owners and other ML principals accountable for failures of ML in \S~\ref{sec:accountability}, through a combination of techniques for attribution and auditing.
    \item We systematize research on techniques for assurance, \ie proactively building confidence that an ML system meets specific security requirements.
    \item We outline the need for mechanisms that manage models throughout their lifecycle, \eg enabling model owners to version usefully. We show how current efforts on machine unlearning provide a {\em partial yet incomplete} response to this need, and that a more comprehensive approach is needed to develop the ability to patch ML systems.
\end{itemize}

We recognize that developing solutions for ML governance is not purely a technical problem because many of the trade-offs that need to be characterized involve a societal component. Given the intended readership of this paper, we put an emphasis on how computer science research can contribute  mechanisms for ML governance, but this assumes that the design of these mechanisms will be informed by societal needs.
Put another way, this SoK leads us to identify opportunities for collaboration with communities beyond computer science, \ie opportunities for interactions that inform the goals of computer science research, but also that would help increase adoption of technical concepts beyond computer science. We thus hope that our manuscript will be of interest to readers beyond the computer security and ML communities.

\section{Background}

\subsection{Machine Learning Primer}

In this paper, we focus on the supervised learning problem in machine learning. We are given a dataset $D = \{(\mathbf{x}_i,\mathbf{y}_i)\}_{i \in [n]}$ of inputs $\mathbf{x}_i$ and labels $\mathbf{y}_i$ (encoded as one-hot vectors) and wish to learn a function $\model$ that can map the inputs $\mathbf{x}_i$ to $\mathbf{y}_i$. For a given task we associate a loss function $\mathcal{L}$ which characterizes the error of a given function $\model$ in predicting the labels of the dataset. A common choice (for classification tasks) for the loss function is cross-entropy, defined as:  
\begin{equation}
\label{eq:cross-entropy}
\centering
\mathcal{L}_{CE}=-\mathbf{y}_i\ln(\model(\mathbf{x}_i))
\end{equation}

The function $\model$, or model, is typically represented as a parameterized function whose parameters, the weights, ideally minimize the loss function $\mathcal{L}$ on dataset $D$. In practice, when a closed-form solution cannot be found, the main process of finding a low error model is done by iteratively tuning weights $\weights$ that parameter the model $\model_{\weights}$ via gradient descent. We typically add a further level of stochasticity to the optimization process by applying the update rule with respect to the loss computed on a batch of data $(\hat{\mathbf{x}},\mathbf{y})$ which is randomly sampled from the dataset $D$. To be specific the update rule is given by:
\begin{equation*}
\weights_{t+1} = \weights_t - \eta \frac{\partial \mathcal{L}}{\partial \weights}|_{\weights_t,\hat{\mathbf{x}}_t}
\end{equation*}
where $\eta$ is a hyper-parameter one tunes.

Once the model is trained, it is deployed to make predictions on points it has not seen during training. To evaluate how well a model will \textit{generalize} to these new points, one typically retains a holdout set of labeled data points to measure the performance of the model upon completing its training. 

\subsection{Operational Life-cycle}
\label{ssec:lifecycle}

Machine learning models have a life cycle. They are trained on collected (or generated data) that is processed and optimized to perform a task which often involves recognizing a pattern in the collected data (training). The model is then used to perform the \textit{learned} task on new data, that is, to infer the said patterns in never-before-seen data. At the high level, we can distinguish 3 operational life-cycle steps which can be further subdivided into smaller steps.

\begin{description}
\item[\textbf{Data preparation 1/2: Collection.}] This is the step that gathers data: complexity and governance difficulties in this process may stem from legal, technical (which data to collect is hard to define), or business issues (there is no incentive to collect that data).
\item[\textbf{Data preparation 2/2: Cleaning.}] Here we  transform a dataset $D$ which has been collected into a dataset $D'$ which is amenable to ML methods. This includes removing data that was improperly collected or performing preliminary feature engineering.
\item[\textbf{Model development 1/2: Training.}] This step represents work done by  experts to design a model architecture, configure its training algorithm, and learn model parameters. From a governance point of view, this step is hard to investigate because it either relies on expert insight or on (deep) learning algorithms often  black-box in nature.
\item[\textbf{Model development 2/2: Testing.}] The model is evaluated with a set of metrics estimated on a holdout set of data before the model is confronted with current production data. This evaluation typically assumes that holdout from the same distribution than training data is available.
\item[\textbf{Model deployment 1/2: Model Serving.}] The model is deployed at scale, often using a different software framework than the one used to train it. Much of our discussion on governance focuses on mitigating risks that manifest themselves through failures at this stage.
\item[\textbf{Model deployment 2/2: Model Retirement.}] This step carries important consequences for governance despite being often omitted. We will discuss how to maintain multiple versions of a model, and what the implications are for choosing to retire (or not) a model. 
\end{description}

\section{Defining ML Governance}

\subsection{Overview}

Conceptually, ML governance can be thought of as a set of solutions for balancing the benefits of applying ML with the risks it creates, often at the level of society. Governance is a prerequisite to build trust between principals involved and affected by an ML development. We describe such principals in detail in \S~\ref{sec:principals}, but for now it is useful to assume that they are three types of principals: the model owner (\ie the company or institution creating and deploying the model), the end user, and a regulator. Governance imposes constraints on how principals interact throughout the lifecycle of an ML application (as covered previously in \S~\ref{ssec:lifecycle}). 

The benefits of an ML pipeline are often application-specific. Typically, these benefits are defined by or at least motivate the model owner, \eg applying ML will generate profit for a company. 
One may think that the model's usefulness (\ie utility) is tied to a performance metric (\eg accuracy) and every other consideration can only decrease it. This is true if we view training as an optimization problem; having additional constraints limits the feasibility set. From an asymptotic\footnote{Given the training dynamics, the constraints may direct the search accurately and improve the solution found as done in LASSO~\cite{tibshirani1996regression}.} statistical learning perspective, any constrained problem can produce an objective that is at most as good as the unconstrained problem.
However, other considerations may be just as important as utility---in particular at steps of the life cycle before or after training. 
For instance, model decisions may demonstrate significant unwanted biases (\eg misclassifying individuals from minorities) and prompt the model owner to reassess their data collection and algorithmic choices. Thus, deploying ML comes with inherent \textbf{risks} to the different principals interacting with the ML pipeline:
users may have concerns with models analyzing their data, regulators may seek to limit potential harm to society  (\eg limit unfair predictions). We detail these risks in \S~\ref{ssec:risks}.

These varied risks, however, do not carry equal meaning and urgency for all principals involved in creating, using, and regulating these models. Because risk definition is inherently tied to a given principal, \textit{\eg} due to their own goals and incentives, and various principal definitions may come into conflict with one another, model governance is best formalized as a \textit{game}. Each principal participating in the game is optimizing for their own set of constrained objectives at any point in time, which improves the principal's trust in the ML pipeline and its other principals. The metrics underlying these objectives may differ from accuracy, and may be worst-case in nature. Thus, many established practices for evaluating ML fall short of providing the guarantees necessary for ML governance.

\subsection{Principals, Incentives, and Capabilities}
\label{sec:principals}

Defining governance requires understanding the various principals in the ecosystem and their capabilities and incentives. We begin enumerating them below. In certain settings these principals may not always be separate entities---a data collector could also be the model builder---in which case such a principal may assume multiple of the roles defined below. %

\begin{enumerate}
\itemsep0em
\item {\bf End User:} This principal is using a system deploying ML. It may be a customer of a company, a patient at a hospital, a citizen asking for government benefits, etc.
\item {\bf Model Owner:} This principal is one with a particular task that can be solved using ML. They communicate their requirements to the model builders, and clearly specifies how this trained model can be accessed (and by whom). 
\item {\bf Model Builder:} This principal is responsible for interpreting the model owner's requirements and designing an ML solution for them. To do so, the builder requires access to representative data from the domain (see data collector below), and an appropriate choice of training algorithm (and approach). The model builder or model owner will then be responsible for deployment.
\item {\bf Data Collector:} This principal is responsible for sourcing the data as per the model owner and builder's requirements so as to ensure that the learnt model performs well and complies with various characteristics (such as ensuring data privacy etc.). 
\item {\bf Adversaries:} The adversary is any entity that wishes to disrupt any phase of the ML life cycle. They may have varying control and access over the ML system. %
\item {\bf Trusted Arbitrator:} This principal is required in situations where (i) auditing of services provided is required (\eg a legal agency or a mandated organization verifying if deployed models are compliant with GDPR), or (ii) there is a requirement for arbitration between two principals (in situations related to ownership resolution). We also refer to this principal as the \textit{regulator}.\footnote{For simplicity, we conflate the entity providing the regulation or standard and the enforcement entity when they in practice differ.}
\end{enumerate}

\subsection{Risks}
\label{ssec:risks}

We highlight three classes of risks relevant to ML governance that have been studied extensively in the literature. We also outline different failure modes inducing these risks (that can occur at different stages of the ML pipeline). Awareness of the risks present provides design goals and considerations that impact the creation, deployment, and retention of the model, and these failure modes will later shape our discussion on accountability and assurance in \S~\ref{sec:accountability} and~\ref{sec:assurance}. Though we do not exhaustively cover all risks due to space constraints, the governance aspects which constitute this paper are generic and can be expected to generalize to other risks. %

\vspace{1mm}
\noindent{\bf 1. Integrity:} Losing integrity (\ie correctness of the system outputs) is the immediate risk that a model owner tries to minimize, for example, via training the most accurate and robust model. The integrity of a model needs to take into account the overall system this model is deployed within. Integrity risk quantifies the lack of alignment between the output of the ML system and what was expected from it. 

The {\em accidental failures} that cause integrity risks stem from the reliance of ML models on data; data used to train models need to possess specific properties and must be (pre-)processed in specific ways to enable effective training (\S~\ref{subsec:data_ownership}). This  ensures that models are resilient to {\em outliers} (data points that are outside the range of those expected)~\cite{gudivada2017data,cortes1995limits}. Another failure mode for integrity is that of {\em concept drifts} (where the distribution of data changes over time)~\cite{lu2018learning,koh2021wilds}. Collectively, both these issues result in the model failing to generalize.

{\em Adversary-induced failures} follow. At training time, adversaries incorporate \textit{data poisons} (which are specifically perturbed data points) which result in poor generalization (\ie indiscriminate poisoning attacks~\cite{munoz2017towards}), or targeted misclassification (\ie targeted poisoning attacks~\cite{munoz2017towards}). Poisoning adversaries can assume that the data collector uses the labels provided with the poisons, or can also specify their own labels~\cite{shafahi2018poison,zhu2019transferable}. The former is known to be ineffective for large parameterized models~\cite{suciu2018does}, while the latter is effective across the spectrum. Poisoning is effective across a range of learning paradigms~\cite{carlini2021poisoning,ma2019policy}, and across different threat models~\cite{fang2020local}. Backdoor attacks use a different strategy; a specific {\em trigger} pattern induces a misclassification, and can be added to {\em any input} whilst generic poisoning ensures misclassification of inputs belonging to a particular class~\cite{liu2017trojaning}.

During inference, adversaries can strategically modify inputs  called \textit{adversarial examples} (whilst ensuring that they are indistinguishable from their benign counterparts). This induces a misclassification~\cite{szegedy2013intriguing,biggio2013evasion}. These evasion attacks are known to be effective across a wide range of hypothesis classes~\cite{szegedy2013intriguing,goodfellow2014explaining,vernekar2019analysis,shafahi2018adversarial,papernot2016limitations},
and learning frameworks~\cite{lin2017tactics,behzadan2017vulnerability,kos2018adversarial}. Initially introduced in vision~\cite{szegedy2013intriguing,biggio2013evasion}, evasion has now been demonstrated for a wide variety of domains including audio~\cite{carlini2018audio,yakura2018robust,qin2019imperceptible} and text~\cite{ebrahimi2017hotflip,alzantot2018generating,garg2020bae,zhao2017generating}. Attacks can also be physically realized~\cite{kurakin2016adversarial,yakura2018robust,Lovisotto2021slap} and constrained such that samples generated for evasion preserve input semantics~\cite{qin2019imperceptible,alzantot2018generating,dreossi2018semantic,jain2019analyzing,qiu2020semanticadv,sheatsley2021robustness}.

\vspace{1mm}
\noindent{\bf 2. Privacy:} Data is fundamental to training. Research has demonstrated how models trained without privacy considerations can leak sensitive information~\cite{shokri2017membership,carlini2019secret}. This has raised concerns about the use of data containing sensitive information and its consequences for individuals and groups alike. Thus, privacy is increasingly becoming a societal norm~\cite{boyd2010privacy}. This is perhaps best illustrated by  legislative work seeking to adapt protective regulations such as the GDPR to the context of ML applications~\cite{gdpr}. Thus, setting aside the value of upholding privacy as a human norm, growing regulation of data usage means that it is in the interest of model owners to respect privacy-preserving practices. Indeed, not mitigating privacy risks in their ML applications could result in a poor reputation for the model owner and compliance complications. 

Additionally, memorization of training data is essential for generalization of ML models~\cite{feldman2020does,Feldman2020what}, and recent work shows encouraging the model to memorize may help its performance and convergence rate~\cite{Katherine2021Deduplicating}. However, memorization could also lead to {\em accidental failures}: a model can memorize sensitive data, such as credit card numbers, inadvertently~\cite{carlini2019secret}. Such secret information may then be leaked at inference time, \eg when the model tries to auto-complete a sentence talking about credit cards. As shown by the authors, the only effective method among those evaluated is differentially private training~\cite{abadi2016deep}, often at the expense of model performance.

There are also known {\em adversary-induced failures}. In a {\em membership inference attack}, an adversary wishes to determine if a given data point was used to train the model. Adversaries  initially assumed  access to confidential information from the black-box model~\cite{shokri2017membership} which was shown sufficient to determine membership. Following this,  research has further improved (and simplified) black-box attacks~\cite{salem2018ml,song2019privacy,jayaraman2020revisiting,truex2019demystifying,yeom2018privacy}, for instance by performing label-only attacks~\cite{li2020label,choquette2021label}. Instead, white-box attacks further leverage information held in model parameters~\cite{sablayrolles2019white,leino2020stolen,song2017machine,nasr2019comprehensive}, or gradient updates~\cite{melis2019exploiting}.

Beyond membership inference, other attacks attempt to reconstruct data. In  attribute inference attacks, the adversary attempts to impute missing values in a partially populated record using access to an ML model~\cite{jia2018attriguard}. Model inversion attacks aim to reconstruct data points that are {\em likely} from the training dataset. In the classification setting, Fredrikson \etal~\cite{fredrikson2015model} exploit confidence information from the prediction vectors (based on the same overfitting assumption made earlier). With a similar intuition, Yang \etal~\cite{yang2019neural} use an autoencoder to reconstruct the input from the confidence vectors, while Zhang \etal~\cite{zhang2020the} use GANs to aid reconstruction.  
More recent attacks (applicable in distributed training regimes, such as federated learning~\cite{konevcny2015federated}\footnote{Federated learning is a paradigm where multiple clients jointly learn a model by training a local model on their private datasets and sharing gradients which are aggregated at a central entity.}) enable successful reconstruction from gradients~\cite{geiping2020inverting,nasr2019comprehensive,melis2019exploiting, hitaj2017deep,wang2019beyond, song2020analyzing}. %

\vspace{1mm}
\noindent{\bf 3. Fairness:} ML models rely on extracting patterns in data, and 
unwanted biases are reflected within historically-collected data~\cite{lum_predict_2016, mattu_machine_nodate}. 
Unmitigated, these biases transfer into algorithmic decisions that are fed back into the decision making process, thus aggravating the situation. Furthermore, algorithmic decisions themselves can also encode new biases that were not initially present in historical data~\cite{bagdasaryan2019differential}; this may be due to poor quality~\cite{suresh_framework_2021}, and even the careless use of fairness constraints.~\cite{cost_of_fairness}

Like integrity and privacy, the challenges of algorithmic fairness can be considered through the lens of attacks and defenses. One can consider, for example, a prejudiced adversary that attempts to sabotage a model performance against a particular sub-group; but this also risks the integrity of the model. But does guaranteeing integrity mean that we have ensured fairness? The answer is no.

Unlike integrity and privacy, fairness risks largely stem from data collection and processing steps. In an ideal world, (i) we would have balanced datasets to train our models on, and (ii) the predictions of models would apply equally to different members of the population which would, in turn, ensure future balancedness of our datasets because model contributions\footnote{Indeed, it is common for model predictions to be incorporated as training data in the future, as can be the case for instance with online learning.} to historical data would be balanced.  But in most areas of critical concern in algorithmic fairness, neither is the case because there is a \textit{de-facto} sampling process often (implicitly) encoded in the processes that produce the data collected~\cite{mitchell_algorithmic_2021}; an essential factor that should thus be considered under data collection. For example, in pre-trials where a judge has to assess the risk of accepting or rejecting bail, the defendants are only being subject to algorithmic predictions because they have been arrested in the first place. Mitchell~\etal~\cite{mitchell_algorithmic_2021} provide a broad overview on various fairness risks associated with three broad categories: (i) {\em data bias:} both statistical--which evolves from lack of representation (\ie selective labels~\cite{selective_labels}), and societal--which indicates a normative mismatch~\cite{suresh_framework_2021}; (ii) {\em model bias:} influence of choice of hypothesis~\cite{chouldechova2017fairer} and even interpretability, and (iii) {\em evaluation:} how today's predictions can alter decision subjects' behaviors, thus requiring new predictions in the future. We will discuss such performative qualities of models in \S~\ref{subsub:performative}.

We quickly note that alleviating fairness risks in the ``ideal'' way above, would inherently be incompatible with controlling privacy and integrity risks, since our samples sizes are limited by the size of the smallest sub-population, which potentially limits the generalization of our models which not only threatens integrity but can also have adverse effects on privacy~\cite{shokri2017membership}.

\subsection{The Governance Games}
\label{sec:game}
\begin{figure*}
    \centering
    \includegraphics[width=0.6\textwidth]{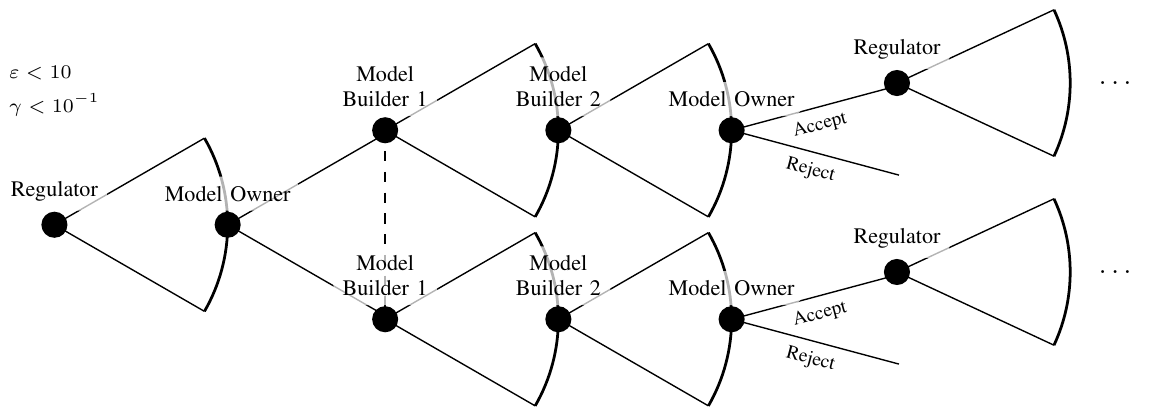}
    \caption{Regulated Model Commission Game in Extensive Form---Strategy sets may be continuous (arcs) or discrete (branches). The dashed line indicates non-singleton information sets: the two decision points of model builder 1 are indistinguishable from model builder 2. The game may be repeated (dots). For brevity, we do not indicate decision spaces of most principals or their pay-offs. 
    }
    \vspace{-5mm}
    \label{fig:game}
\end{figure*}

ML governance follows from the interaction of the principals (\S~\ref{sec:principals}) who are independent decision-makers with goals that may or may not align with each other. For instance, a model owner's desire of achieving robustness may exacerbate leakage of an end user's private information~\cite{song2019membership,song2019privacy} because adversarial training encourages the model to generate predictions that are especially stable for data points that are close to ones in the training set.  We can thus see ML governance as a multi-agent system and analyze dynamics between the principals using standard tools from game theory\cite{shoham_multiagent_2009}. Solutions to this game  affect the degree of trust among principals and towards the application of ML itself. In many governance scenarios, principals act sequentially, after they perceive the effects of other principals' decisions. This naturally leads to a leader-follower, or \textit{Stackelberg} competition structure~\cite{fudenberg_game_1991}.
 
\subsubsection{Stackelberg formulation of learning problems}
\label{subsub:performative}

Work using Stackelberg competitions considers the strategic behavior of a principal (the follower) whose objective is distinct from that of the leader. This improves over the \textit{adversarial assumption} of works that use a zero-sum game formulation (\ie Yao's principle~\cite{yao_principle}) which requires that the follower's objective be exactly the opposite of the leader's. 
The most relevant application of Stackelberg competitions to our discussion is found in strategic classification~\cite{hardt_strategic_2016}. There, the subjects of algorithmic classification are strategic. After observing the decision made, these subjects can adjust their features to achieve a better classification outcome. This is also an example of classification having {\em performative qualities}~\cite{perdomo_performative_2021}: participants adjusting their features causes a distribution shift, which in turn means that the future predictions of the model may be neither fair or optimal w.r.t. the new distribution. In practice, this requires model maintainers to re-train, or employ an online learning setup to preserve their model's integrity and fairness.

Using a similar Stackelberg competition, Perdomo \etal~\cite{perdomo_performative_2021} propose an equilibrium notion called {\em performative stability}. Models that are performative stable produce predictions that would remain optimal under the distribution shifts that result from their own predictions. Similarly, Br\"uckner~\etal~\cite{bruckner_stackelberg_2011} consider spam filtering and analyze an adversarial prediction game between a learner (leader) who tries to classify spam and an adversary (follower) that observes the learner's decisions with the goal of producing less detectable spam.

\subsubsection{Beyond two-player competitions}
As useful as Stackelberg games are, they only cover a subset of interactions that principals may have with each other as part of an ML governance scenario. 
In other scenarios,  several principals may be deciding without observing their opponent actions. For example, a model owner may be unaware of possible future regulatory efforts, an end user may modify its trust assumptions about using ML services, multiple model builder contractors could enter bidding competitions to train a model, etc. This requires a strategic-form game usually captured in the form of a pay-off table~\cite{fudenberg_game_1991}.

However, such a table may be difficult to assemble because interactions can be repeating over time in several stages: principals' expected risks and pay-offs should be adjusted with the perceived effects of their decisions in the future. In such cases, an {\em extensive-form} game which can be represented as a decision tree can capture multi-stage games with observed actions; it shows {(i)} the set of principals, {(ii)} the order of their decisions, {(iii)} their choices at their respective decision points, aka \textit{strategy sets}; {(iv)} their pay-off (or cost) as a function of their own choice as well as the history of the choices made prior to it, {(v)} \textit{information sets}: the information available to each principal at any decision point, and finally {(vi)} the probability distribution over any exogenous event~\cite{fudenberg_game_1991}. 

We believe extensive-form games are a promising avenue to formalize interactions between principals and develop ML governance in a principled way. Figure~\ref{fig:game} illustrates this with a hypothetical ML governance scenario. The regulator would like to ensure fairness and privacy guarantees so it requires that a (pure) DP~\cite{dwork_algorithmic_2013} mechanism be used with a budget of $\varepsilon < 10$. For fairness, the regulator requires demographic parity~\cite{fairml} with at most $\gamma < 10\%$ disparity between sub-populations. The model owner can choose its own specifications for $\varepsilon$ and $\gamma$ and commission two model builders to produce the models. Model builders are now in a bidding competition to achieve the most performant model within specification. They may choose different architectures, source their data from various data collectors (not shown in the figure for brevity), etc. In particular, they get to choose how they allocate their \textit{privacy budget} between these different steps of the ML life cycle~\cite{mironov_renyi_2017}. We note that model builder 2 \textit{does not} observe its rival's decision (hence the non-singleton information set of builder 1 is annotated with dots). Therefore, the sub-game between the two builders is a strategic-form game played simultaneously between the two as no one goes first. Finally, the model owner accepts one model and rejects the other. Model builder pay-offs are assigned at this stage. Next, the regulator can verify achieved fairness and privacy guarantees, and issue proportional fines/incentives for the model owner if they are within specification, or even improve upon it. For instance, privacy-minded regulators could devise fines that scale with the gap between the achieved $\varepsilon$ in practice and the regulatory specified one; or a fairness regulator may propose tax breaks for model owners who have achieved a certifiably fairer prediction model. Since the game is repeated, the regulator can learn from the history of the game and update the regulation (if necessary) for the next stage of the game. 
 
\subsubsection{Solution Concepts} 

Calculating Nash equilibria is an NP-hard problem~\cite{nash_complexity}. As a result, most classical solution concepts for extensive-form games make simplifying assumptions which limit their applicability to governance games. These include {(i)} assuming sub-game perfect equilibria, {(ii)} amenability to backward induction, or {(iii)} agents adopting behavioral strategies~\cite{fudenberg_game_1991}. 
While computing exact strategies may be intractable in the presence of multiple principals or continuous strategy sets, or when the interaction are recurring across a time horizon~\cite{hu_multi-leader-follower_2015, pang_quasi-variational_2005};
approximations such as correlated equilibria (a superset of Nash equilibria)~\cite{fudenberg_game_1991} are easier to find by iterative mechanisms, and have recently seen a flurry of research for their efficient calculation even in the presence of multiple principles and across a time horizon~\cite{meila_multiagent_2021}.

Furthermore, in many realistic scenarios principals may have incomplete information about the game (known as Bayesian games~\cite{fudenberg_game_1991}). For example, in Figure~\ref{fig:game}, model builders may not know the technologies (and therefore, the fairness/privacy guarantees) that their rival is able to achieve.

Recently, new solution concepts have been introduced~\cite{li_solving_2021, fiez20a_implicitlearning, li_endtoend_2020} that tackle these challenges by combining ML with established techniques such as backward induction. 

\begin{tcolorbox}[boxsep=0.5pt,left=0.8pt,right=0.8pt,top=0.5pt,bottom=0.5pt]
\noindent{\textbf{Call to action:}} As discussed in the introduction, research on the limitations of ML often considers a single class of risk. When the interplay between risks is considered, two classes of risk are isolated to study their interplay and integrity is often one of them. For instance, the  ML community has recently started to realize the importance of considering strategic decision makers and subjects which has led to new and practical definitions of robustness\cite{perdomo_performative_2021}. The security community has also long depended on game theoretical studies of potential adversaries~\cite{gametheory_security}. We posit that it is time for the research community to discover the more complex interplay of a realistic governance scenario with {(i)} multiple principals, {(ii)} independent pay-offs and objective functions, {(iii)} allowing for possibly imperfect information. We believe such a multi-faceted approach would alleviate issues that have plagued single-objective approaches, such as choosing hyper-parameter values for parametric definitions of privacy or fairness.
\end{tcolorbox}
\vspace{-3mm}

\section{Establishing Identity in ML Systems}
\label{sec:ownership}

The notion of \textit{identity} is fundamental to computer security~\cite{bishop}. Developing the notion of identities involves first recognizing which principals are involved in a system and second how they interact with this system as it takes actions over objects. Thus, identities are a prerequisite to holding principals accountable for failures of the system (\eg who should be held responsible for the failures of an ML model deployed by a self-driving car?).

Similar to other systems, those that incorporate ML need to integrate access control mechanisms to ensure that principals are authorized to interact with the system. %
We identified a number of principals in \S~\ref{sec:principals}.  For instance, an end user may need to 
authenticate themselves before they are allowed to query an MLaaS API, just like authentication could be necessary to query a database. However, because ML systems introduce new, often implicit, information flows between data and models, they call for specific mechanisms.

In particular, we focus on three areas of interest: the training data, the test data, and the learned model. In the following subsections, we define {\em ownership} as the process of associating a principal with these key components of an ML system. Later, in \S~\ref{sec:accountability}, we focus on how we can hold principals accountable for model failures once identities relevant to data and models have been clearly established through the process of ownership resolution.

\subsection{Data Ownership} 
\label{subsec:data_ownership}

Data is one of the foundational pillars of a {\em functional} ML model, both at training and inference (test) time. 

\vspace{0.5mm}
\noindent{\bf Training Data:} During learning, the training data is used to compute the objective function being optimized to find the model's parameters (recall Equation~\ref{eq:cross-entropy}). Since the existence of training data is a prerequisite to obtaining a model, the entity owning the training data plays an integral role in the governance of the ML system.

Specifically, understanding who collected the data can enable answering various questions regarding (i) {\em representativeness:} if the collector paid attention to ensure that the data collected is not stagnant and captures effects such as domain/concept shifts (where applicable)~\cite{gama2014survey}, (ii) {\em data cleanliness:} if the data is processed for usage, and sanitization techniques (such as the use of robust statistics) are applied~\cite{diakonikolas2019recent}, (where applicable, especially for detecting poisons~\cite{biggio2012poisoning}), (iii) {\em fairness:} if the data collected is representative of the downstream task and various subgroups (as specified by the model owner of \S~\ref{sec:principals})~\cite{holstein2019improving}, and (iv) {\em feature engineering:} if the data was pre-processed correctly for various downstream tasks (such as private/fair learning)~\cite{tramer2020differentially}.

Cryptographic primitives are essential in establishing and protecting data ownership by enforcing confidentiality. Applications of homomorphic encryption~\cite{gentry2009fully} to ML would provide the principal data confidentiality whilst still enabling training. Yet, the use of cryptography introduces computational bottlenecks and makes it difficult to leverage hardware acceleration, often requiring secure hardware~\cite{tramer2018slalom,volos2018graviton}. We discuss the application of homomorphic encryption to ML in more detail in \S~\ref{sec:assurance}. In addition, there has been a shift from centralized learning to distributed learning techniques (such as federated learning~\cite{konevcny2015federated}) to enable co-locating data and computation. Additionally, these learning techniques are {\em assumed} to provide data privacy as the data never leaves the data owner's site~\cite{melis2019exploiting}. Federated learning proposes to have individual data owners compute model updates locally on their data rather than share this data with a central server. The central server then relies on cryptographic primitives (such as secure aggregation~\cite{bonawitz2016practical}) to aggregate updates from different data owners. Other approaches involve using techniques from secure multi-party computation (MPC) to enable training~\cite{wagh2020falcon,wagh2019securenn,mohassel2017secureml,mohassel2018aby3,patra2020blaze,chaudhari2019astra,zheng2019helen,zheng2021cerebro}, but most suffer from poor scalability.

Defining data ownership legally is complex, if only because defining data itself is complex. Under most jurisdictions facts, that is the building blocks of the information recorded directly or indirectly as measurable data, are not protected by copyright law~\cite{scassa2018data}. Hence, the question of data ownership shifts from the protection of individual data records to their collection, arrangement, and derived works. Collections of data and derived works from data may be unprotected by copyright law depending on a number of conditions such as (i) under the merge doctrine~\cite{2007new,2013banxcorp}, if their exposition is inherently tied to the facts and data they represent, or (ii) if the arrangement is not considered original enough, as evaluated in court, \eg a phone book alphabetical order is not protected. For ML, this may entail that if a data collection pipeline being deployed is deemed not sufficiently novel, the collected dataset may not be protected by copyright law.  Observe that copyright on a dataset would not prevent someone from using individual data points but rather significantly copying the entire dataset. Depending on the jurisdiction, the status of ownership over personal information and the existence of a personal data ownership right is debated~\cite{scassa2018data}. Canada leans towards only providing access but not ownership of personal information whereas in the EU the existence of a personal data ownership right has been argued.

\noindent{\bf Test Data:} At inference, the model is deployed to predict unseen data. Unlike training data, test data does not influence the model permanently (\ie it does not impact the values of the model parameters). That said, test data can still be responsible for model behavior that may make ML governance more difficult to achieve. For instance, test data may contain adversarial examples~\cite{goodfellow2014explaining} that affect the integrity of model predictions or sensitive data that should not be revealed to the model owner directly. Thus, like for training data, both technical and legal means are required to establish and protect test data ownership.

As discussed earlier, cryptography can address certain aspects of test data ownership as well. For instance, homomorphic encryption enables a model to perform inference without having a cleartext view of the input point~\cite{juvekar2018gazelle,mishra2020delphi,lehmkuhl2021muse}. The same can be achieved using techniques from MPC. This is helpful in scenarios where proprietary medical records can not be shared between institutions~\cite{alvarez2020secure,kaissis2021end,DBLP:journals/corr/abs-2107-10230}. Here again, moving from a central to a distributed setup for ML can help foster trust for the data owners~\cite{choquette2021capc}. %

\subsection{Model Ownership} 
\label{subsec:model_ownership}

Model ownership is often a broad term used to refer to the ownership of the model's {\em sensitive parameters} that were obtained after the (computationally intensive) training process. Defining ownership is necessitated by the existence of various threats that infringe the confidentiality of the model, and the need to be able to hold principals that own ML models accountable for the failures of their models. %

\vspace{0.5mm}
\noindent{\bf What is model extraction?} Attackers may target the confidentiality of ML models with a model extraction attack for three reasons: (i) data collection and labeling is time-consuming and expensive~\cite{halevy2009unreasonable}, (ii) the training process is computationally expensive (especially in many deep learning applications)~\cite{strubell2019energy}, and (iii) knowledge of the model can serve as reconnaissance to mount a follow-up attack against the model in the white-box threat model rather than black-box. Models that are proprietary could be released through insider threats~\cite{10.1145/1413140.1413158}, and models that are accessible via an interface can be reverse engineered~\cite{lowd2005adversarial}. If the latter is a query interface where principals can pose inference queries to models, then it is formalized as model extraction~\cite{tramer2016stealing}: the objective of the adversary is to succeed with {\em as few queries} as possible to avoid detection. By repeatedly posing (inference) queries to the trained model, the adversary recovers the model's {\em functionality}. But functionality is an abstract concept; Jagielski \etal~\cite{matthew2020high} provide a detailed taxonomy of what exactly an adversary can recover through this reverse engineering process, ranging from the exact model parameters to replicating output behaviors for specific input data points. 

Extraction is demonstrated in various domains, such as images~\cite{matthew2020high}, text~\cite{krishna2019thieves}, and for simple hypothesis classes~\cite{varun2020exploring}, and more complicated ones such as DNNs~\cite{carlini2020cryptanalytic, rolnick2020reverse}; assuming grey-box access~\cite{pmlr-v139-zanella-beguelin21a}, and more strict black-box regimes~\cite{milli2019model}. Some recent attacks do not pose inference queries, but assume access to hardware side-channels~\cite{batina2019csi,zhu2021hermes}; while such attacks recover the model, they are not model extraction attacks in the traditional sense. 

\vspace{0.5mm}
\noindent{\bf But how does the adversary generate queries?} Assuming that the adversary has access to the dataset used to train a model, or the distribution from which it was sampled from, is a strong assumption to make. To alleviate these concerns, Truong \etal~\cite{truong2021datafree} propose a refinement to the work of Kariyappa \etal~\cite{kariyappa2021maze}; they propose a {\em data-free} extraction method based on disagreement between the victim (\ie trained model) and model being trained by the adversary. A query synthesis active learning-based approach was used by Chandrasekaran \etal~\cite{varun2020exploring} for extracting simple hypothesis classes; this provides asymptotic bounds on the number of queries needed. The same authors also argue that pool-based active learning strategies can be used when the adversary does have access to a pool of unlabelled data, and Jagielski \etal~\cite{matthew2020high} propose using semi-supervised learning techniques such as MixMatch~\cite{berthelot2019mixmatch} to practically instantiate extraction. %

\vspace{0.5mm}
\noindent{\bf Why is it impossible to prevent model stealing?} Various defenses have been proposed to prevent the adversary from learning enough information to enable extraction~\cite{juuti2019prada}. These range from reducing precision of the outputs~\cite{tramer2016stealing}, to more sophisticated approaches such as randomized model selection~\cite{kariyappa2020protecting,alabdulmohsin2014adding}, adding noise in responses~\cite{kariyappa2020defending}, protecting data points near decision boundaries by differential privacy~\cite{zheng2019bdpl}, or perturb outputs to mislead gradient-based optimization algorithms (\eg SGD)~\cite{tribhuvanesh2020prediction}.
However, such approaches are rendered ineffective with an adaptive extraction strategy~\cite{varun2020exploring}, or introduce utility degradation. Chandrasekaran \etal~\cite{varun2020exploring} state the inevitability of extraction. The authors argue that any utilitarian model (\ie one that is reasonably accurate) leaks information which enables extraction. %

\vspace{0.5mm}
\noindent{\bf How can one prove ownership?} To prove ownership, the process of {\em watermarking} was considered. Digital watermarking has been used to covertly embed a marker into digital media to later prove ownership infringements~\cite{petitcolas1999information}. In the context of watermarking ML models, the model owner encodes a specific query response pair into the training procedure (analogous to embedding a marker), and can verify if a model is indeed their own on observing such behaviours at the time of inference~\cite{adi2018turning}. This process is fundamentally similar to the concept of {\em backdooring}~\cite{gu2017badnets}, without any adversarial intent (with the additional requirement that watermarks are non-forgeable). However, recent work~\cite{jia2020entangled} demonstrates that watermarking is not resilient to model extraction; Jia \etal~\cite{jia2020entangled} propose an approach that remediates this issue, but with a nominal performance penalty. 

All forms of watermarking rely on encoding specific secret information {\em into} the training algorithm. Recent research suggests that there exists various forms of secret information \textit{already} available to the true model owner, obtained {\em during} the process of training. This includes the order in which data is sampled, intermediate checkpoints, hyperparameter values, the intermediate values of model weights, etc. One proposal to prove model ownership relies on the honest model owner logging such information during the training procedure. Since DNN training is a stochastic process, the intermediary states obtained during training are {\em hard to guess} without access to the secret information aforementioned. If a trusted third party can reproduce these states (within some acceptable tolerance threshold) with the help of the honest principal's secret information, then the honest principal's ownership of the model is validated. This is the premise of the {\em proof-of-learning} approach to enable ownership~\cite{jia2021proof}. 

\begin{tcolorbox}[boxsep=0.5pt,left=0.8pt,right=0.8pt,top=0.5pt,bottom=0.5pt]
\noindent{\bf Call to action:}\todo{put that in a grey box?} First, notice that claiming model ownership and differentiating between two models are two sides of the same coin. It is unclear if existing metrics (such as the distributional distance between model parameters or predictions) capture the {\em closeness}, and research is needed on this front. Second, most techniques to detect extraction rely on minor changes to the training procedure. More research is needed to understand if intrinsic properties of the ML model  can serve as its fingerprint (ergo enabling ownership resolution).
\end{tcolorbox}

\subsection{Is model ownership separate from data ownership?}

Maini \etal proposed dataset inference that validates whether a model was trained using a specific training set (or training-set distribution)~\cite{maini2021dataset}. This is related to model ownership because as long as only the model owner has access to the (proprietary) training data, the model owner can prove ownership. However, in other cases (such as training using public data), this technique cannot prove ownership.

Recall that from a legal angle, collections and derived works from data may at times be unprotected by copyright law. Because ML is known to memorize~\cite{feldman2020does}, models might not be protected by copyright law unless the model builder can show that their implementation non-trivially included significant original work. Similarly, content synthesized by generative models~\cite{zhang2020attribution}, and  language models such as GPT-3~\cite{brown2020language} could be, as derived work, respectively protected or in violation of copyright law~\cite{scassa2018data}. In Canada, but not necessarily in other jurisdictions like the European Union, copyright generally requires at least one human author---thus excluding automated art generation processes from copyright protection~\cite{scassa2018data}. 
Given that instances of sufficiently original (and/or difficult) derived work have been allowed copyright~\cite{scassa2018data}, if a DNN satisfies such a condition that would allow for the distinction between data ownership and model.

\section{Holding ML Principals Accountable}
\label{sec:accountability}

Identities, as established in \S~\ref{sec:ownership}, serve multiple purposes~\cite{bishop}; one of them involves enabling{\em accountability}. In conjunction with traditional access control mechanisms, the notion of identity provided by data and model ownership in the context of ML allows us to establish the responsibility of principals interacting with an ML system. This is defined as accountability, and is different from the {\em process} of ascertaining accountability which we define as {\em attribution}. Indeed, accountability is essential when ML models produce {\em undesirable} behaviors (\ie failures) such as the ones discussed in \S~\ref{ssec:risks}. Some of these failure modes will be known and need to be attributed to a principal (we outline these earlier in \S~\ref{sec:principals}). This is similar to how malware analysis will seek to identify whether a malware originated from a state actor or not. Principals held responsible could be the model owner, but also end users (\eg if they are attempting to steal a model, or to have it produce an incorrect prediction with an adversarial example). Other failure modes of the model may be initially unknown but uncovered by external scrutiny of an ML pipeline, \eg by a regulator through an auditing process discussed later in this section.

\subsection{Attribution of Known ML Failures}
\label{subsec:attribution}

Recall that we define attribution as the process used to identify principals responsible for model failures. Here, we discuss detection of the failures discussed earlier in \S~\ref{ssec:risks}. This is a precursory requirement for attribution.

\vspace{0.5mm}
\noindent{\bf 1. Outliers:} A considerable amount of research investigates how to identify when DNNs are predicting on out-of-distribution (OOD) samples. The model's confidence for different outputs can be made more uniform on OOD samples~\cite{meinke2019towards,lee2017training,hein2019relu,hendrycks2018deep,lakshminarayanan2016simple,vyas2018out}, or the model can be trained explicitly to assign a confidence score which can be used to tell how likely the input was out of distribution~\cite{devries2018learning,vernekar2019analysis,malinin2018predictive}. Other methods have also been introduced to distinguish the outputs of DNNs on OOD and in-distribution samples~\cite{liang2017enhancing,lee2018simple,ren2019likelihood,serra2019input,hsu2020generalized}, including approaches based on inspecting the model's internal representations, training set, or performing statistical testing~\cite{raghuram2021general,sastry2019detecting,jiang2018trust,papernot2018deep,jha2019attribution}. These approaches could be applied to hold any principal querying the model on OOD samples accountable.

\noindent{\bf 2. Concept Drift:} The model builder may attempt to detect drifts in the training data distribution to hold the data collector accountable. Lu \etal \cite{lu2018learning} summarize past research into three broad themes: (i) error rate based detection (a model's deteriorating performance serves as an indicator of shifts)~\cite{gama2004learning,gama2006learning,baena2006early,frias2014online,liu2017fuzzy,xu2017dynamic,huang2006extreme,ross2012exponentially,nishida2007detecting,bifet2007learning,bifet2009adaptive,bifet2009improving,bifet2009new,gomes2017adaptive}, (ii) data distribution based detection (which try and discern differences in the distribution of previous and new inputs)~\cite{kifer2004detecting,dasu2006information,song2007statistical,qahtan2015pca,bu2017incremental,liu2017regional}, and (iii) multiple hypothesis test drift detection (which differ from the previous two by utilizing more than one test to discern drift)~\cite{alippi2012just,wang2015concept,zhang2017three,du2015selective,maciel2015lightweight,alippi2016hierarchical,raza2015ewma}. Note that the two latter approaches could also be used by the data collector to hold potential downstream data contributors accountable.%

\vspace{0.5mm}
\noindent{\bf 3. Data Poisoning:} Attribution of poisoning can be done either at the data collection or model training stages. To the best of our knowledge, the only known techniques belonging to the former category utilize insight from outlier/anomaly detection~\cite{liu2017neural,Chen2017TargetedBA}. For the latter, research focuses on finding relationships between poisoned data and special behaviors in the representation space (or activations of hidden layers) ~\cite{tran2018spectral,chen2019detecting,liu2019abs}, investigate the gradients~\cite{Hong2020on}, or observe changes in the learned decision boundaries~\cite{wang2019neural,chen2019deepinspect}. 

\vspace{0.5mm}
\noindent{\bf 4. Adversarial Examples:} %
To distinguish adversarial examples from legitimate inputs, detection techniques often first pre-process the inputs. This includes replacing the input by its internal model representations~\cite{lu2017safetynet, wong2018provable}, transforming the inputs~\cite{xu2018feature, roth2019the}, applying dimensionality reduction techniques~\cite{hendrycks2017early, bhagoji2017dimensionality}.
After pre-processing, a classifier or statistical tests distinguish distributions of adversarial and legitimate inputs~\cite{grosse2017statistical,hendrycks2017early,roth2019the,raghuram2021general}. %
However, only a small fraction of the proposed detection techniques are evaluated against an adaptive attacker~\cite{tramer2020adaptive}, which may prevent robust attribution of adversarial examples and jeopardize accountability.

\vspace{0.5mm}
\noindent{\bf 5. Model Extraction:} To attribute queries contributing to a model extraction attack, Juuti \etal~\cite{juuti2019prada} compare the distribution of distances between queries. However, if an adversary leverages public data (which is in-distribution), the proposed detection method fails. It is unknown if other techniques can be devised for this in-distribution querying regime as well.

\vspace{0.5mm}
\noindent{\bf 6. Privacy Attacks:} There currently exist no mechanisms to identify if privacy attacks (such as membership inference or data reconstruction) are occurring at inference time. This is due to the nature of the attack: adversaries query the model with (potentially) in-distribution data and observe the outputs. 

\begin{tcolorbox}[boxsep=0.5pt,left=0.8pt,right=0.8pt,top=1pt,bottom=1pt]
\noindent{\bf Call to action:}\todo{grey box?} Discovering approaches that can detect and attribute (adaptive)  membership inference while it is being performed remains an open problem.
\end{tcolorbox}

\vspace{0.5mm}
\noindent{\bf 7. Synthetic Content:} Generative models have revolutionized content generation~\cite{karras2019stylebased,zhang2019selfattgan,park2019semantic}; while this is useful in many scenarios (such as private data release~\cite{abay2019privacy}), it also introduces a myriad of problems. For instance, content produced by generative adversarial networks~\cite{gan_goodfellow} is being actively used in disinformation campaigns~\cite{bbc20chinastylegan}. A number of approaches were proposed in order to {\em detect} such fake content~\cite{mccloskey2018detecting,Li2020identification,zhang2019detecting,marra2018gans} (including large-scale competitions~\cite{facebook2020deepfakechallenge}). Attributing synthetic content to a particular generative model is difficult because models are often capable of producing~\cite{zhang2020attribution} and encoding~\cite{abdal2019image2style} any arbitrary content and allow for encoding of arbitrary contents. A promising approach is to study model components explicitly, \eg upsampling components which produce spectral aliasing~\cite{zhang2019detecting}.

\subsection{Auditing to Identify New Model Failures}

Our discussion thus far has focused on techniques devised to detect failure models that are well known and understood. An alternative involves auditing mechanisms. To this end, we first define ML auditing and present how ML logging can support it. We then discuss storage, privacy, and security (open) problems that result.

\vspace{0.5mm}
\noindent{\bf What is ML auditing?} Bishop defines an audit as a formal examination by a (third) party to determine a chain of events and actors~\cite{bishop}. In the case of ML, this definition must be extended to include the verification of the {\em claimed properties} of a model, such as its lack of bias or presence of adversarial robustness.
As noted in \S~\ref{sec:principals}, the different logical entities may, in practice, refer to the same principal. Hence, we call an audit \textit{internal (resp. external)} when the model owner and auditor are the same entity (resp. independent entities). Observe that an internal audit is a proactive action similar to a white-box assessment, and may rely on model logs \ie data records collected on the model and its inputs. On the contrary, an external audit is similar to a black-box assessment, and may only rely on model querying when the model owner cannot be trusted. Prior literature tends to conflate audit with external audit. This leads to a pessimistic view of ML audit as a non feasible task. This distinction helps us introduce alternate auditing methods including logs which are promising but completely unexplored. This apparently simple distinction provides deeper insight: (i) during internal audits, the regulators have white-box access to model internals, code, and logs; this enables easier verification of purported functionality as opposed to external audits where regulators only observe the functionality with no internal knowledge, and (ii) leveraging internal logs may alleviate technical limitations of an external audit \eg the current lack of guarantees with regards to model robustness when the auditor can only query the model. %

For instance, the model deletion code (\ie internal audit) of SISA unlearning~\cite{bourtoule2019machine} can be reviewed for correctness \ie one only needs to check if a model is being trained without the data-point to be unlearned, while external verification (\ie external audit) of unlearning (through query access) remains challenging (because approaches like membership inference introduce a large number of false positives). For private learning, mechanisms like DP-SGD~\cite{abadi2016deep} come with a privacy accountant that trivially tracks expenditure, but the privacy of the models can also be audited through computationally expensive means post-hoc~\cite{jagielski2020auditing,nasr2021adversary} In the context of algorithmic fairness, model owners are being asked to conduct fairness audits~\cite{whitehouse, kim_auditing_2017}. Despite the well-intentioned tools that have arrived as a result~\cite{2018aequitas}, given poor incentives for model owners, we are now faced with the risk of fake fairness audit reports~\cite{faking_fairness}. Prominently, Aivodji~\etal demonstrates how using an interpretable proxy allows model owners to evade accurate fairness audits~\cite{aivodji2019fairwashing}.

Much like system and application logs are well-known as a fundamental tool in incident investigations in non-ML settings~\cite{bishop}, a model should never be separated from its logs.

\vspace{0.5mm}
\noindent{\bf Why is logging hard for ML?} In general, knowing what and when to log is non-trivial and highly use-case dependent. Additionally, log storage space is finite, and overhead induced by logging is challenging in many other domains~\cite{bates_trustworthy_nodate}. Logging in ML differs from non-ML applications because of (i) ML pipeline control-path explosion dependent on external events, the data, and randomness which increases the content required to be logged, (ii) the different ML-specific steps, like training and testing, of the model life cycle which require logging very diverse and step specific information, and (iii) there is a need, parallel to that in system and application logs, to focus on both identifying principals responsible for a specific high-dimensional input but also, specifically for ML, tracking certain model properties (required for accountability and attribution respectively). %
Consequently, the ML logged information may greatly differ from traditional event logs used in application, network, and system security. For example, the proof (or log, albeit it is not referred as such) created in the work of Jia \etal~\cite{jia2021proof} is created with the explicit goal of proving model (parameter) ownership and requires information about data used to train the model, randomness used in the training process (which is hard to precisely capture), and tracks the change in model parameters as training progresses.

\vspace{0.5mm}
\noindent{\bf Maintaining log confidentiality.} Lastly, when designing an ML auditing system based on logs, the privacy and security of that system itself should be taken into account. 
Here again, ML introduces specific issues. First, since logs may be accessed by auditors, they represent a risk and must be sanitized~\cite{bishop}. As illustrated in other sections, this sanitization is not trivial and in the case of logging is further complicated by its stream nature. This (important) requirement makes ML logging hard in practice. Second, adversaries may try to simultaneously pose threats to the ML model and log analysis system (if it is ML-based). For example, Travers \etal~\cite{travers2021exploitability} propose evasion attacks for audio-based ML models whilst ensuring that intermediary steps (which can be considered as logs) are within acceptable levels.

\begin{tcolorbox}[boxsep=0.5pt,left=0.8pt,right=0.8pt,top=1pt,bottom=1pt]
\noindent{\bf Call to action:}\todo{grey bubble?} Despite the upstart of a number of industry ML logging tools and practices (both for the model and underlying system performance)~\cite{noauthor_whylogs_2021}, only little academic work has delved into the challenges of ML logging. 
\end{tcolorbox}

\section{Assurance in ML}
\label{sec:assurance}

In addition to being transparent about failures resulting from the ML system's use, which they can be held responsible for as shown previously in \S~\ref{sec:accountability}, ML principals will need to establish trust in their application of ML techniques. Bishop distinguishes trust, which is the often ill-defined belief that a system is secure, from \textit{assurance} which is a set of techniques by which one builds confidence from evidence that a system meets its security requirements~\cite{bishop}. Here, by security we refer to the broader mitigation of risks identified in \S~\ref{ssec:risks}. One well-known example of such techniques is formal methods for verification, which have started to emerge for ML systems~\cite{raghunathan2018certified,cohen2019certified,sommer2020towards,singh2019abstract,singh2018fast}. We discuss these techniques and others, following the operational life-cycle described in \S~\ref{ssec:lifecycle}

\subsection{Assurance during Data Preparation}
\label{subsec:assurance_data}

Assurance in the context of data preparation ensures that models learnt satisfy specific properties (\eg privacy, robustness, fairness, etc.). Assurance can be provided through data sanitization (\eg removal of data poisons)~\cite{gudivada2017data,cortes1995limits} and checks for concept drift~\cite{lu2018learning}, bias alleviation~\cite{wang2020revise,zhang2018mitigating}, and subgroup representation~\cite{buolamwini2017gender,selwood2002politics}. To the best of our knowledge, there has been no work that provides formal guarantees about a dataset providing the aforementioned properties. It is also unclear how such guarantees would be provided. An alternative approach to providing assurance involves providing documentation showing the processes used during data collection met certain criteria that are designed to provide assurance; Gebru~\etal~\cite{gebru2018datasheets} proposed datasheets for datasets, a generic documentation scheme meant to be used to verify the data collector employed practices sufficient to alleviate known failures, and other work have further expanded on (or proposed similar measures)~\cite{holland2018dataset,miceli2021documenting,costa2020mt,hanley2020ethical}.

\subsection{Assurance for Training}

As in the case of prior components of the life cycle, assurance can either be obtained {\em post-hoc} (\ie after training completes), or be obtained during training. Recall that assurance is required among various axes, which we describe next.

\vspace{1mm}
\noindent{\bf Integrity:} In the context of evasion attacks, robustness is {\em provably} provided using two main strategies. The first is adversarial training~\cite{madry2017towards} and its variants~\cite{shafahi2019adversarial,xie2019intriguing,xie2020smooth,tramer2019adversarial,andriushchenko2020understanding,shafahi2020universal,wang2019convergence,wong2020fast}. Such strategies are formalized such that assurance is encoded into the training algorithm and an auditor would have to inspect the algorithm itself to understand what properties were provided. An alternative approach involves providing a certificate of robustness~\cite{cohen2019certified, salman2019provably} which provides a certificate indicating that the trained model is robust to perturbations within a pre-defined $p-$norm radius. To the best of our knowledge, Steinhardt \etal~\cite{Steinhardt2017certified} proposed the first certified data poisoning defense by proving an upper bound on the model's loss in the presence of an adversary. However, as they pointed out, their methods have to run on clean data, which is often impractical. Ma \etal~\cite{Ma2019DataPA} propose DP as a solution against poisoning and provide theoretical bounds on the number of poisons tolerable by such models. Weber \etal~\cite{Weber2020RABPR} and Levine \etal~\cite{Levine2021deep} both propose newer certified defenses against data poisoning; the former uses techniques similar to randomized smoothing~\cite{cohen2019certified} and the latter uses collective knowledge from an ensemble of models to prove robustness. Finally, integrity of the computation and model parameter values can also be verified by a mechanism that provides a proof associated with the particular run of gradient descent which produced the model parameters, as suggested by Jia \etal~\cite{jia2021proof}. 

\vspace{1mm}
\noindent{\bf Privacy:} Several approaches can be taken to preserve data privacy. The de-facto approach is to utilize {\em DP learning}. Proposed initially by Chaudhuri \etal~\cite{chaudhuri2011differentially}, DP learning has evolved from the objective perturbation approach proposed by the authors to more sophisticated techniques which involve adding noise during optimization~\cite{abadi2016deep}, or during aggregating votes from disjoint learners~\cite{papernot2018scalable}. These newer approaches have tighter forms (and analyses) of privacy accounting (\ie calculating $\varepsilon$). In the decentralized setting, approaches such as CaPC~\cite{choquette2021capc} provide assurances about data confidentiality and privacy. Recent research~\cite{jagielski2020auditing, nasr2021adversary} demonstrates that the privacy accounting during training is not tight (in practical scenarios), and could benefit from external audits.

\noindent{\bf Randomness:} Unsurprisingly, providing assurance that sources of randomness in the process of learning are cryptographically secure is essential. Research demonstrates that, if manipulated, the order of the batches served to a model during training can itself result in a lack of generalization guarantees from the training algorithm, resulting in a lack of integrity for the model parameters output by this training algorithm~\cite{shumailov2021manipulating}. Furthermore, randomness in data sampling is fundamental to amplification results demonstrated for DP learning~\cite{abadi2016deep,erlingsson2020encode,kairouz2021practical}. Providing assurances on this matter is an open problem.

\noindent{\bf Fairness:} For individual fairness, \cite{ruoss_learning_2020} produces a formal certification by ensuring the representations of similar points are kept close in the latent space. For subgroup fairness definitions, however, 
the primary focus
has been in designing unfairness mitigation techniques despite data bias issues.
Considering a core ML task (often classification) as an example, these techniques can be categorized into {(i)} pre-processing, where algorithms re-weight, obfuscate, transform, and/or remove samples to satisfy a notion of fairness~\cite{wang2020revise,feldman2015certifying,calmon2017optimized,kamiran2012data,zemel2013learning}; {(ii)} in-processing, where additional constraints are added to the optimization procedure to result in fairer models~\cite{zhang2018mitigating,kamishima2012fairness,celis2019classification,ruoss2020learning}; or {(iii)} post-processing, where the output is changed to match a desired distribution~\cite{hardt2016equality,pleiss2017fairness,kamiran2012decision}. Other methods have been developed to empirically assure the fairness of a model by thoroughly documenting its performance on different sub-populations~\cite{mitchell2019model,yang2018nutritional}.

Despite improving fairness, an important limitation of these techniques is that they do not strictly provide formal fairness guarantees. To obtain assurance of fairness during the training of a model, algorithms that provably result in fair models must be implemented. Again, initial work in this vein provides guarantees through
pre-pocessing~\cite{feldman2015certifying} and  in-processing~\cite{celis2019classification,zemel2013learning,ruoss2020learning}. The most well-known formal guarantees for prediction fairness is provided under a metric-based notion of individual fairness~\cite{individual_fairness}: Rothblum~\etal introduce a PAC-like relaxation of the said notion to prove asymptotic fairness guarantees for the individual~\cite{pacf}.

\begin{tcolorbox}[boxsep=0.5pt,left=0.8pt,right=0.8pt,top=1pt,bottom=1pt]
\noindent{\bf Call to action:}
The algorithmic fairness community has recently recognized the importance of considering the broader ``algorithmic system''~\cite{studying_up}, to consider social actors and decision making context. While we agree with the message, we posit that we need to go beyond studying fairness---or any single class of risk on its own---and consider the complex but realistic interplay of the principals in the broader context of ML governance. Given the importance of data in all such systems, a concrete step is to research fairer, more accurate, and non-privacy-invasive data collection practices. 
\end{tcolorbox}
\vspace{-2mm}

\subsection{Assurance at Deployment}
\label{subsec:assurance_deployment}

When a model is deployed as a service, there are mainly two entities possibly interacting with the model: the service provider who publishes the model, and the end users querying the model. In this section, we discuss assurance provided by the service providers to the end user.

To build end user confidence, service providers may aim to show two features of their service: (i) integrity, that is the computation is performed correctly and the prediction is indeed made by the model being advertised (rather than for instance a simpler but cheaper model), and (ii) maintain confidentiality of the query made by the users which can contain sensitive information. For example, hospitals may query online healthcare ML models to diagnose patients' diseases, where 
the queries could contain the patients' (private/confidential) health condition. %

\vspace{1mm}
\noindent{\bf Integrity:} End users would require assurance regarding the correctness of the prediction generated by the model. The field of verified computations~\cite{walfish2015verifying} is designed to allow a principal to verify the correctness of outsourced computation. The main issue with using {\em off-the-shelf} techniques from verified computations literature on DNN inference is the large overhead incurred. To solve this issue, building on Thaler \etal's~\cite{thaler2013time} work towards efficient verified computing, Ghodsi \etal~\cite{ghodsi2017safety} introduce SafetyNets which applies verified computing techniques to inference operations efficiently. 
Note that verifying the integrity of computation is fundamentally different from approaches that assure confidence of (or trust in) the prediction~\cite{jiang2018trust} %

\vspace{1mm}
\noindent{\bf Confidentiality:} Another concern of using MLaaS is regarding the confidentiality of the data used to query the ML models. To this end, several approaches rely on homomorphic encryption~\cite{dowlin2016cryptonets,liu2017oblivious,juvekar2018gazelle,mishra2020delphi,reagen2021cheetah}, or use MPC techniques for inference~\cite{wagh2020falcon,choquette2021capc}. Both these approaches rely on cryptographic techniques which are provably correct.

\section{Life Cycle Management of ML Systems}
\label{sec:life-cycle-management}

The techniques presented thus far do not preclude ML models from needing modifications: training from scratch to remove certain pathological behaviors learned by the model may be expensive~\cite{bourtoule2019machine} and leveraging backups will not solve issues which are embedded in the backup model's weights. The process of patching an ML system, a form of maintenance, is thus the last component of our approach to governance. In the following, we outline what these different processes would look like for ML systems.

\subsection{Patching ML Systems}

\noindent{\bf 1. Unlearning:} After deployment, the model owner may lose access to part of the data used to train the model. This may be due to detecting data poisons, or outliers, or for legal reasons. For example, this may be the case when the data collector raises privacy concerns and invokes the right-to-be-forgotten~\cite{bertram2019five,mantelero2013eu}. The process of obtaining a model without the influence of said data-point(s) is referred to as machine unlearning~\cite{cao2015towards}.

Various forms of unlearning have been proposed which offer different guarantees. For example, retraining a model from scratch would give a concrete guarantee to the end-user the unlearned model is now independent of the data the user revoked access to. However, this comes with large (training) costs. One could modify the model's architecture and training pipeline to reduce the retraining cost~\cite{bourtoule2019machine}, but to some the cost may still be too high. Alternatively, the approach of Sekhari \etal~\cite{sekhari2021remember} states that a model with strong privacy guarantees (\ie DP-privacy) does not need to be modified to forget a user as it does not leak user-specific information in the training dataset, and thus unlearning requires no additional computational resources. However, private models are often not utilitarian~\cite{tramer2020differentially,bagdasaryan2019differential}. To find a middle ground between computational costs and performance, an alternative approach is to define an unlearning rule (for example adding a hessian-vector product) that modifies a model such that the resultant model would be similar to a model that had not trained on the users' data~\cite{guo2019certified,graves2020amnesiac,golatkar2020eternal,baumhauer2020machine}.

Unlearning highlights the importance of modeling interactions between ML principals and the interplay between different risks raised (\S~\ref{sec:game}). For example, a user's decision to revoke access to their information can itself leak information about them having been in the training dataset if it is made adaptively~\cite{zanella2020analyzing,gupta2021adaptive,chen2020machine}. %

\begin{tcolorbox}[boxsep=0.5pt,left=0.8pt,right=0.8pt,top=0.8pt,bottom=0.8pt]
\noindent{\bf Call to action:} Regulators will play an important role in deciding what should be forgotten (and to what degree) for the various scenarios when parts of the training dataset must be removed. It could be that any unlearning method that mitigates privacy leakage would suffice for dealing with users revoking access to their data. In contrast, it might be that for data poisoning (where there could be grave security risks) the model owner would need to completely retrain or give some guarantee of changing the weights sufficiently to remove the threat. Determining this is subject to future research. 
\end{tcolorbox}

\noindent{\bf 2. Fine-tuning:} One may also want to additionally learn new model behaviors. This can be achieved by fine-tuning models. Fine-tuning is needed in two scenarios: in (i) online learning~\cite{fontenla2013online}, \ie more and more data becomes available after the model is trained so it is preferable to fine-tune the model on more data to improve its performance, and (ii) transfer learning~\cite{tan2018survey}, where one wants to fine-tune a trained model to help a different but related task, \eg classifying traffic signs used in another country.
If the model's capacity remains fixed or training data previously used is not repeatedly used to train the model, one must take care in avoiding catastrophic forgetting: by learning a new behavior, the model may inadvertently perform poorly on data points it has previously analyzed~\cite{French1999Catastrophic}.

\subsection{Retiring ML Systems}

Following our discussion of unlearning, it is easy to see that one may consider retiring an ML model altogether when a large portion of the training data is requested to be unlearned (and the remaining data is insufficient to obtain a performant model). Apart from unlearning, there may be other reasons for a model owner to choose to retire an ML pipeline. This includes ethical concerns where retiring the data itself may help prevent further ethical concerns, \eg as discussed by Peng \etal~\cite{peng2021mitigating}. Regulators will here again play a crucial role, and may decide to develop frameworks that encourage ML principals to retire their data and models when appropriate.

\section{Discussion}

Our systematization led us to develop a framework for ML governance. Through the concepts of ownership, accountability, assurance, and life cycle management, we foster trust between principals involved in an ML application. Our game formulation proposes to explicitly search for equilibria that balance the benefits of ML with its risks. Space constraints prevented us from discussing certain aspects that remain relevant to achieving governance. For instance, deploying ML involves computer systems that support ML but are not captured in our analysis; such computer systems are implicitly assume to form the trusted computing base. Hence, our solutions for governance largely rely on the security of these systems themselves. Beyond this aspect, we now outline avenues for future work to strengthen our understanding of governance.

\vspace{1mm}
\noindent{\bf Risk Interplay:} Through this work, we have delved into understanding various facets of risk associated with ML. The game formulation explains the interaction between various principals and how this leads to a constant recalibration of risk. Yet, we pointed out how there remains a limited understanding of the interplay between the various risks themselves. %
Limited research has been conducted to analyze the interplay of the various facets of risk in unison; research however has been conducted to understand this interaction in a pair-wise manner (\ie relationship between utility and robustness~\cite{dohmatob2019generalized}, or privacy and fairness~\cite{chang2020privacy}), but this paints an incomplete picture as discussed in \S~\ref{sec:game}.

\vspace{1mm}
\noindent{\bf Inter-principal Trust:} An important goal of governance is to improve trust between ML principals. Hence, documenting clearly the limitations of ML systems is of paramount importance to communicate clearly when a satisfactory approach to governance cannot be found. For this reason, we believe that approaches analogous to the one taken by dataset cards~\cite{gebru2018datasheets} will complement well our work on governance. This also relates to research on explainability which seeks to enable end users to understand the logic behind learning algorithms. There are two lines of work here: (i) feature-level, and (ii) concept-level attribution. The former attributes every feature (\eg pixels in an image) to a prediction~\cite{ribeiro2016lime,kindermans2018learning, Kapishnikov2021guided}. The latter attributes higher-level concepts to a prediction~\cite{kim2018tcav,zhou2018interpretable,yeh2020completeness}.

\vspace{1mm}
\noindent{\bf Interpretability:} The techniques discussed in \S~\ref{sec:accountability} arm principals in identifying various issues. However, they may not understand {\em why} these issues occur and consequently may have a difficult time resolving them. Ensuring that the learning and prediction process of many ML algorithms is interpretable (particularly to non-experts) is paramount to enable more widespread and trustworthy usage. 

\vspace{1mm}
\noindent{\bf Parameterization:} Furthermore, the assurance provided is often parameterized. For example, the assurance provided by DP learning is the expenditure of privacy ($\varepsilon$). However, interpreting these parameters is non-trivial for the average user. It is also cumbersome for the user to specify their requirements (for example, privacy requirements) to model builders for the same reason. Parameterized definitions are not easily accessible, and this further impedes governance. Additionally, recent research~\cite{kaptchuk2021need} suggests that user expectations for privacy do not match what is provided. Thus, this difference between mental models and reality can also impede governance. While similar to the earlier discussion on interpretability, we wish to highlight that parameterization is more focused on providing principals knowledge to bootstrap the training process, while interpretability provides tools to better understand its outcomes.

\vspace{1mm}
\noindent{\bf Trust in Regulators:} We implicitly assumed that the regulator principal is trustworthy and seeks to preserve the society from ML risks. This may not always be the case, for instance in settings where ML is used by authoritarian regimes. Additionally, trusted regulators may not be privy to sensitive information in reality, as such forms of data transfer are strictly governed by laws~\cite{2016re}. 

\vspace{1mm}
\noindent{\bf Legal Challenges:} While we have taken care to highlight certain legal challenges raised by ML governance, there currently lacks interdisciplinary work on law and ML. Studying the technical problem in isolation is unsatisfactory because without a match between law and techniques, techniques will not see adoption. For instance, the technical community focuses on DP but legal frameworks focus on anonymization~\cite{gdpr}. This gap results in a lack of adoption of advances made to privately learn from data. Sometimes the inverse problem holds: while laws against disparate impact have long been in effect; it has been fairly recently that we have seen their adoption by the ML community~\cite{feldman2015certifying}.

\newpage
{
\scriptsize
\bibliographystyle{IEEEtran}
\bibliography{references}
}

\end{document}